\documentclass[twocolumn,showpacs,showkeys,preprintnumbers,superscriptaddress,
amsmath,floatfix,amssymb,secnumarabic,english]{revtex4}
\usepackage[hypertex]{hyperref}
\usepackage{graphicx}
\newcommand\nn{\nonumber}

\newcommand{\const}{\mbox{const.}}

\newcommand{\Tr}{{\mathrm{\rm Tr}}}

\newcommand{\GeV}{\mbox{GeV}}

\newcommand\ba{\begin{eqnarray}}
\newcommand\ea{\end{eqnarray}}

\newcommand{\bc}{\begin{center}} 
\newcommand{\ec}{\end{center}}   
\newcommand{\vecc}[1]{\mbox{\boldmath $#1$}}
\newcommand{\be}{\begin{equation}}
\newcommand{\ee}{\end{equation}}

\begin{document}
\def\bs{\boldsymbol}

\title{On the Physical Meaning of Sachs Form Factors and on the Violation\\
of the Dipole Dependence of $\bs{G_E}$ and $\bs{G_M}$ on $\bs{Q^2}$ }

\author{M.~V.~Galynskii}
\email{galynski@dragon.bas-net.by}
\affiliation{Joint Institute for Power and Nuclear Research - Sosny, BAS, 
 220109 Minsk, Belarus}

\author{E.~A.~Kuraev}
\email{kuraev@theor.jinr.ru}
\affiliation{ 
Joint Institute for Nuclear Research, Dubna,
Moscow region, 141980 Russia }
\begin{abstract}
In this work we discuss questions related to the interpretation of unexpected results of measurements
of the proton form factors ratio $G_{E}/G_{M}$ in the high-precision double polarization
experiments done in JLab in the region of $0.5 \leq Q^2 \leq 8.5 \, \GeV ^2$.
For this purpose, in the case of the hard scattering mechanism we calculated
(in the leading approximation) the matrix elements of the proton current $J^{\pm \delta, \delta }_{p}$
for the full set of spin combinations corresponding to the number of the spin-flipped quarks, which contribute
to the proton transition without spin-flip ($J^{\delta, \delta }_{p}$) and with the spin-flip
($J^{-\delta, \delta }_{p}$). This set is: (0,1), (0,3), (2,1), (2,3), where the first number
in parentheses is the number of the spin-flipped quarks, which contribute to the $J^{\delta, \delta }_{p}$,
and the second one is the number of the spin-flipped quarks which contribute
to the $J^{-\delta, \delta }_{p}$. For the sets of (0,1) and (2,3), we found that the ratio
$G_{E}/G_{M} \sim 1$, and the form factors $G_E$ and $G_M$ behave for the set of (0,1) as
$G_E, G_M \sim 1/Q^6$, and for the set of (2,3) as $G_E,G_M \sim 1/Q^4$.
At the same time the set of (0,1) is realized for $\tau \ll 1$, and the set (2,3)
for $\tau \gg 1$ ($\tau=Q^2/4m^2$). This allows us to suppose that: 1) at the lower boundary
of the experimental measurements of the ratio $G_{E}/G_{M}$ not dipole dependence appears
but the law of $G_E, G_M \sim 1/Q^6$; 2) the conditions for the observation of the dipole dependence
in the experiments has not yet been achieved; 3) since for quarks $J^{\delta, \delta }_{q} \sim 1$
and $J^{-\delta, \delta }_{q} \sim  \sqrt{\tau}$, then the dipole dependence is realized
when $\tau \gg 1$ in the case when the quark transitions with spin-flip are dominate;
4) the law of the linear decrease of $G_{E}/G_{M}$ at $\tau < 1$ is due to additional
contributions to the $J^{\delta, \delta }_{p}$ by spin-flip transitions of two quarks
and an additional contribution to $J^{-\delta, \delta }_{p}$ by spin-flip transitions
of three quarks, in this case their relative contributions are small.
\end{abstract}

\maketitle
\section{Introduction}
\label{sec:introduction}

Experiments aimed at studying the proton form factors, the electric ($G_E$) and magnetic ($G_M$) ones,
which are frequently referred to as the Sachs form factors, have been performed since the mid-1950s
\cite{Rosen,Hof58} by using elastic electron-proton scattering. In the case of unpolarized
electrons and protons, all experimental data on the behavior of the proton form factors were
obtained by using the Rosenbluth formula \cite{Rosen} for the differential cross section
for the reaction $ep \to ep$; that is,
\ba
\label{Ros}
\frac{d\sigma} {d\Omega_e}=\frac{\alpha^2E_2\cos^2(\theta_e/2)}{4E_1^3\sin^4(\theta_e/2)}
\frac{1}{1+\tau}\left(G_E^2 +\frac{\tau}{\varepsilon}G_M ^2\right).
\ea

Here, $\tau=Q^2/4m^2$, $Q^2=-q^2=4E_1 E_2\sin^2(\theta_e/2)$ is the square of the momentum
transfer to the proton and $m$ is the proton mass; $E_1$, $E_2$ and $\theta_e$ are, respectively,
the initial-electron energy, the final-electron energy, and the electron scattering angle
in the rest frame of the initial proton; the quantity $\varepsilon$ is the
degree of virtual photon linear polarization, $\varepsilon^{-1}=1+2(1+\tau)\tan^2(\theta_e/2)$; and $\alpha=1/137$
is the fine-structure constant. Expression (\ref{Ros}) was obtained in the approximation
of one-photon exchange. In deriving it, the electron mass was set to zero.
With the aid of Rosenbluth's technique, it was found that the experimental dependences of $G_E$ and
$G_M$ on $Q^2$ are well described up to 10 GeV$^2$ by the dipole-approximation expression
\ba
G_E =G_M/\mu=G_D(Q^2) \equiv (1+Q^2/0.71)^{-2}\,,
\label{eq:GMSL}
\ea
where $\mu$ is the proton magnetic moment ($\mu=2.79$).

In \cite{Rekalo68}, Akhiezer and Rekalo proposed a method for measuring the ratio of the Sachs form factors. Their
method relies on the phenomenon of polarization transfer from the longitudinally polarized initial
electron to the final proton. They showed that the ratio of the degrees of longitudinal ($P_l$)
and transverse ($P_t$) polarizations of the scattered proton has the form
\ba
\frac{P_l}{P_t}=-\frac{G_M}{G_E}\frac{E_1+E_2}{2m} \tan \frac{\theta_e}{2}\,.
\label{AxRek}
\ea
Precision experiments based on employing Eq. (\ref{AxRek}) were performed at JLab and were
reported in \cite{Jones00,Gayou02}. They showed that, in the range of $0.5<Q^2<5.6$ GeV$^2$,
there was a linear decrease in the ratio $R = \mu G_{E}/G_{M}$ with increasing $Q^2$:
\begin{equation}
R =1-0.13\,(Q^2-0.04)\,,
\label{linfit}
\end{equation}
which indicates that $G_E$ falls faster than $G_M$. In the non-relativistic limit,
this fact could be interpreted as indicating that the spatial distributions of charge
and magnetization currents in the proton are definitely different.
This is at contradicts with data obtained with the aid of Rosenbluth's technique. According to those data,
$G_E$ and $G_M$ approximately follow the dipole form up to the value of $Q^2 \simeq 10$ GeV$^2$;
concurrently, the approximate equality $R \approx 1$ must hold. Repeated, more precise,
measurements of the ratio $R$ using the polarization transfer method \cite{Puckett10,Meziane11,Puckett12}
and by Rosenbluth's method \cite{Qattan} only confirmed this contradiction, showing that the magnetic
form factor did not differ within the errors from its counterpart obtained within Rosenbluth's
technique and that the electric form factor fall short of the respective value in accordance with
Eq. (\ref{linfit}).

In order to resolve this contradiction, it was assumed that the discrepancy in question may be
caused by disregarding, in the respective analysis, the contribution of two-photon exchange. There
appeared a large number of articles devoted to this problem (see \cite{Brodsky2005,Perdrisat2007};
see also the review article of Arrington et al. \cite{Arrington2011} and references therein).
At the present time, three experiments aimed at studying the contribution of two-photon exchange are known.
These are an experiment at the VEPP-3 storage ring in Novosibirsk, the OLYMPUS experiment at the
DORIS accelerator at DESY in Hamburg (Germany), and the EG5 CLAS experiment at JLab (USA).

In \cite{GKB2008}, we proposed a method for determining the Sachs form factors in the process
$ep \to ep$ on the basis of measuring cross sections for spin-flip and non-spin-flip transitions for protons.

The objective of the present study is to show that the fundamental physical meaning of the form factors
$G_E$ and $G_M$ is associated with their factorization in the matrix elements of the proton current
that correspond to non-spin-flip and spin-flip transitions for protons. It is precisely this
circumstance that explains the appearance of the squares of the Sachs form factors in
Rosenbluth's cross section.

Yet another objective of this study is to show that the mechanism of one-photon exchange is sufficient
for explaining the results of the polarization experiment at JLab. Namely we state that
in these experiments in the region of transfer momenta $0.5 \leq Q^2 \leq 8.5 ~ \GeV^2$ the conditions
for realization of the dipole dependence of proton form factors $G_E$ and $G_M$ on $Q^2$
are still not fulfilled.
Below we show that near the lower bound of this region ($Q^2\approx 1 ~ \GeV^2 $) the unit
value of the form factors ratio is provided by behavior of kind $G_{E}\approx G_{M} \sim 1/Q^6$.
It correspond to the case when only one quark in the proton has spin-flip transition.
Dipole dependence holds at high values of $Q^2$ when the number of the flipped-spin quarks
which contribute to the proton transfer without and with spin-flip is a maximal possible.

\section{On the physical Meaning of the Sachs Form Factors}

Rosenbluth's cross section in the rest frame of the primary proton (\ref{Ros})
has a compact form owing to the decomposition of $G^2_E$ and $G^2_M$. In text-books on particle physics,
it is shown that the physical meaning of the form factors $G_E$ and $G_M$ is that,
in the Breit frame of the initial and the final proton, they
describe the distributions of the proton charge and magnetic moment, respectively; this means that, in
the Breit frame, the matrix elements of the proton current for non-spin-flip and spin-flip transitions for the
proton are expressed in terms of $G_E$ and $G_M$, respectively. Moreover, the Sachs form factors are
advantageous in view of the simplicity of expression (\ref{Ros}).

The question of whether there is any physical meaning behind the decomposition of $G^2_E$
and $G^2_M$ in Rosenbluth's cross section was not raised and not discussed  either in textbooks
or in scientific literature. Nevertheless, it was shown
many years ago in the article of Sikach \cite{Sik84} that the form factors $G_E$ and $G_M$
factorize in the diagonal spin basis (DSB) even at the level of amplitudes in calculating
(in an arbitrary reference frame) the matrix elements of the proton current in the cases
of non-spin-flip and spin-flip transitions for the proton.

\subsection{Diagonal Spin Basis}

In DSB, the spin 4-vectors $s_{1}$ and $s_{2}$ of fermions with 4-momenta $q_{1}$
(before the interaction) and $q_{2}$ (after it) have the form \cite{Sik84}
\ba
s_{1} = - \; \frac { (v_{1} v_{2}) v_{1} - v_{2}} {\sqrt{(v_{1}v_{2} )^{2} - 1 }} \; , \; \;
s_{2} =  \frac { ( v_{1} v_{2})v_{2} - v_{1}} {\sqrt{ ( v_{1}v_{2} )^{2} - 1 }} \; \; ,
\label{DSB}
\ea
where $v_{1}=q_{1}/m$ and $v_{2}=q_{2}/m$. Obviously, the spin 4-vectors in (\ref{DSB})
satisfy ordinary conditions -- that is, $s_{1} q_{1}=s_{2} q_{2}=0$ and $s_{1}^{2}=s_{2}^{2}=-1$ --
and are invariant under the transformations of the little group of Lorentz group
(little Wigner group \cite{Wigner}) $L_{q_1 q_2}$ common to particles with 4-momenta $q_{1}$
and $q_{2}$: $L_{q_1 q_2} q_1 =q_1$ and $L_{q_1 q_2} q_2 =q_2$.
We note that this group is isomorphic to the one-parameter subgroup of the rotational group $SO(3)$
with an axis whose direction is determined by the three-dimensional vector \cite{FIF70,GL}
\ba
\vecc a = \vecc q_{1}/q_{10} - \vecc q_{2}/q_{20}  \; .
\label{veca}
\ea
For the two particles in question, the spin projections onto the direction specified by the
vector in Eq. (\ref{veca}) simultaneously have specific values \cite{FIF70,GL}, and the concept of non-spin-flip
and spin-flip transitions acquires an absolute physical meaning.

The vector $\vecc a$ in Eq. (\ref{veca}) is the difference of two three-dimensional vector,
and the geometric image of the difference of two 3-vectors is a diagonal of the parallelogram spanned by these
two vectors. This is the reason why the term ``DSB'' was introduced by academician F.I. Fedorov.

Let us consider the realization of DSB in the initial proton rest frame, where
$q_1=(q_{10},\vecc q_1)=(m, \vecc 0)$. In this case for the vector $\vecc a$ in Eq.(\ref{veca})
we have: $\vecc a=\vecc n_2=\vecc q_2/|\vecc q_2|$;
that is, the direction of final proton motion is a common direction onto which one projects
the spins in question. Consequently, the polarization state of the final proton
is a helicity state, while the spin 4-vectors $s_{1}$ and $s_{2}$ in (\ref{DSB}) have the form
\ba
\label{s1s2}
s_1=(0,\vecc n_2 ), \, s_2= (|\vecc v_2|, v_{20}\, \vecc {n_2}),  
\ea
that is, the axes of the spin projections $\vecc c_{1}$ and $\vecc c_{2}$ coincide with
the direction of final-proton motion: $\vecc c_{1} =\vecc c_{2}=\vecc n_2$.

The Breit frame, where $\vecc q_2=-\vecc q_1$, is a particular case of the DSB.

\subsection{Spin Operators and Calculation of Amplitudes for QED Processes in DSB}

In DSB (\ref{DSB}), the spin projection operators $\sigma_{1}$ and $\sigma_{2}$ for
the initial and final Dirac particles with 4-momenta $q_{1}$ and $q_{2}$ coincide,
as well as the respective raising and lowering spin operators $\sigma_{1}^{\pm\delta}$
and $\sigma_{2}^{\pm\delta}$, by virtue of the realization of the little Lorentz group $L_{q_1q_2}$
in DSB and have the form \cite{GS89,GS98}
\ba
\label{spop6a}
&&\sigma = \sigma_{1} = \sigma_{2} =\gamma^{5} \hat{s_{1}} \hat{v_{1}} =
\gamma^{5} \hat{s_{2}} \hat{v_{2}}  = \gamma^{5} \hat{b}_{0} \hat{b}_{3},\nn  \\
&&\sigma^{\pm\delta} = \sigma_{1}^{\pm\delta} =\sigma_{2}^{\pm\delta} = -i/2 \gamma^{5} \hat{b}_{\pm\delta},
b_{\pm\delta} = b_{1} \pm i \delta b_{2},  \label{spop6b}\\
&&\sigma u^{\delta}(q_{i}) = \delta u^{\delta}(q_{i}), \sigma^{\pm\delta} u^{\mp\delta}(q_{i})
= u^{\pm\delta}(q_{i}), \delta = \pm 1,\label{spop6c} \nn
\ea
where $u^{\delta}(q_{i})=u^{\delta}(q_{i}, s_{i})$ are the bispinors of the initial
and final states of the particles in DSB; $\hat{s}_1=(s_1)_{\mu}\gamma^{\mu}$, $\gamma^5,
\gamma^{\mu}$ are the Dirac matrixes.

In expressions (\ref{spop6a}), an orthonormalized basis of vectors $b_{A} \,( A=0, 1, 2, 3)$,
\ba
&&(b_1)_{ \mu} = \varepsilon_{\mu \nu \kappa \sigma}b_0^{\nu}b_3^{\kappa}b_2^{\sigma},\;
(b_{2})_{\mu} = \varepsilon_{\mu \nu \kappa \sigma}q_1^{\nu}q_2^{\,\kappa}r^{\sigma}/\rho , \nn \\
&&b_{3} = q_-/ \sqrt{-q_- ^2}, \;b_0=q_+/\sqrt{q_+^2} \; ,
\label{OBV}
\ea
%
was used to construct the respective spin operators. Here, $q_-=q_2-q_1, \,q_+=q_2+q_1$,
$\varepsilon_{\mu\nu\kappa\sigma}$ is the Levi-Civita tensor ($\varepsilon_{0123}=-1$),
$r$ is the participant-particle 4-momentum differing from $q_{1}$ and $q_{2}$, and $\rho$
is determined from the normalization conditions $b_{1}^{2} = b_{2}^{2} =
b_{3}^{2}=-b_{0}^{2}=-1$.

The matrix elements for QED processes have the form
\ba
M^{\pm\delta,\delta} = \overline {u}^{\pm \delta}(q_{2}) Q u^{\delta}(q_{1}) \; ,
\label{MQED}
\ea
where $Q$ is the interaction operator and $u^{\delta}(q_{1})$ and $u^{\pm \delta}(q_{2})$
are the bispinors of, respectively, the initial and the final state.

In the approach that we use (see Appendix \ref{DSBmethod}), the calculation of matrix elements
(amplitudes) that have the form (\ref{MQED}) and which correspond to the fermion transition
from the initial state $u^{\delta}(q_{1})$ to the final state $u^{\pm \delta}(q_{2})$ reduces
to evaluating the trace of the product of Dirac operators \cite{GL,GS89,GS98}; that is,
\be
 M^{\pm\delta,\delta} = Tr (P_{21}^{\pm\delta,\delta} Q ) \; , \; P_{21}^ {\pm\delta,\delta} =
u^{\delta}(q_{1}) \; \overline {u}^{~\pm \delta}(q_{2}) \; . \\
\label{MQED2}
\ee
The explicit form of the operators $P_{21}^{\pm\delta,\delta}$ in DSB that correspond to non-spin-flip
($P_{21}^{\delta,\delta}$) and spin-flip ($P_{21}^{-\delta,\delta}$) transitions was
obtained in \cite{GS89,GS98} and is given by
\ba
&&P_{21}^{\delta,\delta} = ( \hat q_{1} + m ) \, \hat b_{\delta} \,  \hat b_{0}
\; \hat b_{\delta}^{\ast} /4 \; , \label{P21pp2} ~~~~\\
&&P_{21}^{-\delta,\delta} = \delta (\hat q_{1} + m) \; \hat b_{\delta} \; \hat b_{3} /2\;,
\label{P21pm2}~~~
\ea
where $b_{\delta}^{\ast}=b_{-\delta} = b_{1} - i \delta b_{2}$ and
$b_{\delta} b^{\ast}_{\delta}=-2$.

\subsection{Amplitudes of the Proton Current in DSB}

In the Born approximation, the matrix element corresponding to the process of elastic
electron - proton scattering,
\ba
e(p_1)+p\,(q_1,s_1) \to e(p_{2}) + p\,(q_2,s_2)\,,
\label{EPEP}
\ea
has the form
\ba
\label{Mepep}
M_{ep\to ep} = \overline{u}(p_{2}) \gamma^{\mu} u(p_{1}) \cdot \overline{u}(q_2)
\Gamma_{\mu}(q^{2}) u(q_1) \;  \frac {1} {q^{2}} \; ,\\
\Gamma_{\mu}(q^{2}) = F_{1} \gamma_{\mu} + \frac{F_{2}} {4M}
(\hat q \gamma_{\mu} - \gamma_{\mu} \hat q \; ) \; ,~~~~~~~
\label{Gamuepep}
\ea
where $u(p_{i})$ and $u(q_{i})$ are the bispinors of, respectively, the electrons and protons
with 4-omenta $p_{i}$ and $q_{i}$ [accordingly, we have $p_{i}^{2} = m_e^{2}$, $q_{i}^{2} = m^{2}$,
$\overline{u}(p_{i})u(p_{i})=2m_e$, and $\overline{u}(q_{i}) u(q_{i})= 2m \,(i = 1,2)$]; $F_{1}$
and $F_{2}$ are, respectively, the Dirac and Pauli form factors; $q = q_{2}-q_{1}$ is the
4-momentum transfer to the proton; and $s_1$ and $s_2$ are the polarization 4-vectors of,
respectively, the initial and final protons.

The matrix elements of the proton current that correspond to non-spin-flip and spin-flip
transitions for the proton are given by
\ba
( J^{\pm\delta ,\delta }_{p} )_{\mu} = \overline{u}^{\pm \delta }(q_2)
\Gamma_{\mu}(q^{2}) u^{\delta }(q_1) \; .
\label {Jprot0}
\ea
With the aid of Eqs. (\ref{MQED2}) -- (\ref{P21pm2}), we can readily show
that the matrix elements of the proton current in (\ref{Jprot0})
that are calculated in DSB (\ref{DSB}) have the form \cite{Sik84,GS98}
\ba
\label {Jepep-pp}
&&( J^{\delta,\delta }_{p} )_{\mu} = 2 m \,G_{E} ( b_{0} )_{\mu} \, , \\
\label {Jepep-pm}
&&( J^{-\delta,\delta }_{p} )_{\mu}=- 2  m \,\delta \sqrt{\tau} G_{M} (b_{\delta })_{\mu}\, ,\\
&&G_{E} = F_{1} +\frac{q^2}{4m^2}\, F_{2} \, , \; G_{M} = F_{1} + F_{2} \;,
\label {FFSep}
\ea
where $G_E$ and $G_M$ are the Sachs form factors and the quantities $\tau=Q^2/4m^2$,
$Q^2=-q^2$, $q=q_-=q_2-q_1$, $b_0$, and $b_{\delta}$ were defined above.

We note that the amplitudes of the proton current in (\ref{Jepep-pp}) and
(\ref{Jepep-pm}) satisfy the conditions of gauge invariance since, by virtue
of the definitions of the 4-vectors $b_0$ and $b_{\delta}$, the scalar products
$b_0 q$ and $b_{\delta}q$ are equal to zero. Further, the matrix element
$(J^{\delta,\delta }_{p} )_{\mu}$ of the proton current in (\ref{Jepep-pp}) for
the non-spin-flip transition for the proton is expressed in terms of the 4-vector $b_0$.
This matrix element corresponds to the exchange of a virtual photon that has a scalar
polarization ($b_0^2=1$) and which therefore cannot carry away a spin moment.
At the same time, the matrix element $(J^{-\delta ,\delta }_{p} )_{\mu}$ in (\ref{Jepep-pm})
for the spin-flip transition for the proton is expressed in terms of the complex
4-vector $b_{\delta}$. It corresponds to the exchange of a virtual photon having a circular
polarization vector ($b_{\delta}^2=0, b_{\delta}b^{\ast}_{\delta}=-2$) and carrying away
a spin moment, with the result that there occurs proton spin-flip. Thus, our analysis
of expressions (\ref{Jepep-pp}) and (\ref{Jepep-pm}) obtained for the matrix elements in question leads
to the conclusion that these expressions are fully adequate to the physical picture of the phenomena being
considered.
It follows that the electric and magnetic form factors $G_E$ and $G_M$ acquire a fundamental physical
meaning owing to their factorization in the matrix elements of the proton current for non-spin-flip and
spin-flip transitions for the proton. It is precisely because of the factorization
of $G_E$ and $G_M$ in the amplitudes in Eqs. (\ref{Jepep-pp}) and (\ref{Jepep-pm}) that
the contributions to Rosenbluth's cross section for non-spin-flip and spin-flip
transitions for the proton are controlled by the terms containing $G_E^2$ and $G_M^2$, respectively.

In the case of pointlike particles having a mass $m_q$, the amplitudes for
their currents have the form
\ba
\label{Jtpp}
&&~( J^{\delta ,\delta }_{q} )_{\mu} = 2 \,m_q \,( b_{0} )_{\mu} \, ,\\
&& ( J^{-\delta ,\delta }_{q} )_{\mu}=-2\,m_q\, \delta^{} \sqrt{\tau_q}\,(b_{\delta })_{\mu}\, ,
\tau_q=Q_q^2/4m_q^2 \;.
\label{Jtpm}
\ea
In the ultrarelativistic massless case, only spin-flip transitions [see Eqs.
(\ref{Jepep-pm}) and (\ref{Jtpm})] contribute to the cross section for the process
being considered, since the amplitudes in (\ref{Jepep-pp}) and (\ref{Jtpp}) vanish.
At first glance, this conclusion contradicts the well-known fact that,
in the ultrarelativistic limit, only processes in which the particle helicity
is conserved survive at high energy; that is, only amplitudes corresponding to
non-helicity-flip transitions do not vanish in the massless limit.
Such processes are frequently referred to as non-spin-flip processes. However,
this terminology is quite uncertain since the particles involved have different
directions of motion before and after the interaction event.
Moreover, it is erroneous since, in non-helicity-flip processes, the spins of
the particles are in fact flipped at high energies. There is no contradiction here
since, in DSB, the initial state for ultrarelativistic particles is a helicity state,
while the final state has a negative helicity \cite{GS98} (see Eqs. (\ref{ga5tau}) and (\ref{tauga5})),
with the result that
\ba
M^{-\delta,\delta}=M^{-(-\lambda),\lambda}=M^{\lambda,\lambda},\,
M^{\delta,\delta}=M^{-\lambda,\lambda}=0 \,.
\label{mspir}
\ea

We note that, in addition to the representation in (\ref{Gamuepep}) for $\Gamma_{\mu}(q^2)$, the following
equivalent representation is used in the literature for  this quantity:
\begin{eqnarray}
\Gamma_{\mu}(q^2) =G_M \gamma_{\mu}- \frac{(q_1+q_2)_{\mu}}{2m} \, F_2 \;.
\label{Gamu2}
\end{eqnarray}
On the basis of explicit form (\ref{Gamuepep}) and (\ref{Gamu2}) for $\Gamma_{\mu}(q^2)$,
in the literature it is likely just starting with the paper of Lepage and Brodsky
\cite{Brodsky1980} stated that the Dirac (Pauli) form factor $F_1$ ($F_2$)
corresponds to helicity-non-flip (helicity-flip) transitions of the proton, respectively.
In fact, it is the form factor $G_E$ ($G_M$) rather than $F_2$ ($F_1$)
[see Eq. (\ref{Jepep-pp}), (\ref{Jepep-pm}), (\ref{mspir})] that is responsible
for helicity-flip (helicity-non-flip) transitions at high $q_1$ and $q_2$.

We also note that in the literature sometimes there is no clear understanding of the physical
meaning of the quantity $\varepsilon$ in formula (\ref{Ros}).
So in \cite{Jones00,Puckett12,Arrington2011,Qattan,Andi} written that
the quantity $\varepsilon$ is a degree of the longitudinal polarization of the virtual photon.
In fact $\varepsilon$ is the degree of linear polarization of the virtual photon
(see \cite{Rekalo68,GLev} and Appendix \ref{Linpol}).

\section{ON THE VIOLATION OF THE DIPOLE CHARACTER OF THE $\bs{Q^2}$ DEPENDENCE
OF $\bs{G_E}$ AND $\bs{G_M}$}

Since $|b_0|=1$ and $|b_\delta b^*_\delta|=2$, the $Q^2$ dependence
of the absolute values of the matrix elements of the
proton (\ref{Jprot0}) and pointlike-particle ($J^{\pm\delta,\delta}_{q}$) currents
can readily be obtained from Eqs. (\ref{Jepep-pp}), (\ref{Jepep-pm}), (\ref{Jtpp}), and
(\ref{Jtpm}). The results are
\ba
\label {Qzavprot}
&&    J^{\delta ,\delta }_{p} \sim \,2\,m \;G_{E} , J^{-\delta ,\delta }_{p} \sim \, 2m \sqrt{\tau}\; G_{M}\,,\\
&&    J^{\delta ,\delta }_{q}  \sim  2 \,m_q \,, \,
J^{-\delta ,\delta }_{q} \sim 2 m_q \sqrt{\tau_q} \;.
\label {Qzavquark}
\ea
We note that the factorization of $2m$ and $2m_q$ in expressions
(\ref{Jepep-pp}), (\ref{Jepep-pm}), (\ref{Jtpp}), (\ref{Jtpm}), (\ref{Qzavprot}),
and (\ref{Qzavquark}) is due to normalizing the particle bispinors by the condition
$\bar u_i u_i=2m_i$. In performing further calculations, it is more convenient
to employ the normalization condition $\bar u_i u_i=1$. Instead of expressions
(\ref{Qzavprot}) and (\ref{Qzavquark}), we will then use the expressions
\ba
\label {Qprot}
&&    J^{\delta ,\delta }_{p} \sim \,G_{E} , J^{-\delta ,\delta }_{p} \sim \, \sqrt{\tau}\; G_{M}\,,\\
&&    J^{\delta ,\delta }_{q}  \sim  1 \,, \;\;
J^{-\delta ,\delta }_{q} \sim  \sqrt{\tau_q} \;.
\label {Qquark}
\ea
Relations (\ref{Qprot}) and (\ref{Qquark}) make it possible to show how
there arise the dipole dependence of $G_E$ and $G_M$ on $Q^2$
and its violations observed in the aforementioned JLab experiments.

It is commonly accepted in frames of QCD that in the region $Q^2\gg 1 \, \GeV^2$ the hard part
(kernel) of the proton current (\ref{Jprot0}) can be presented as a summ of contributions
where proton is replaced by a set of three almost on mass shell quarks
\cite{Chernyak1977,Chernyak1984}.
Each of the relevant Feynman amplitudes contains two gluon Green functions, of order
of magnitude $1/Q^2$ and, besides two quark Green functions of order $1/Q$.
Below, we will employ that 
the respective absolute values of the proton current matrix elements $J_p^{\pm\delta ,\delta}$ (\ref{Qprot})
are the product of three point-quark-current amplitudes $J_{q}^{\pm\delta,\delta}$ (\ref{Qquark})
divided by $Q^6$
\ba
J_p^{\pm\delta ,\delta} \sim J_{q}^{\pm\delta,\delta } J_{q}^{\pm\delta,\delta }J_{q}^{\pm\delta,\delta }/Q^6\,.
\label{pQCD}
\ea
It is necessary to note that representation (\ref{pQCD}) is valid in the region $Q^2\gg 1 \, \GeV^2$.
Below we will suppose the masses of quarks $m_q$ to be equal to $1/3$ of the proton mass $m$ and the fraction
of the transfer momenta of them to be equal. So we have
\ba
\tau_q=\tau\,.
\label{tau tau0}
\ea
There are two possibilities for a proton non-spin-flip transition:
(i) none of the three quarks undergoes a spin-flip transition, and (ii)
two quarks undergo a spin-flip transition, while the third does not.
We denote the number of such ways as $n^{\delta,\delta}_q=[0,2]$ in
accordance with the number of quarks involved in a spin-flip process (none or two).

Proton spin-flip can also proceed in two ways: (i) one quark undergoes a spin-flip
transition, while the other two do not, and (ii) all three quarks undergo
a spin-flip transition. We denote the number of such ways by $n^{-\delta,\delta}_q=[1,3]$
in accordance with the number of quarks involved in a spin-flip process (one or three).
Thus, there are in all four combinations to be considered:
\ba
n^{\delta,\delta}_q\times n^{-\delta,\delta}_q=(0,1)+(0,3)+(2,1)+(2,3)\,.
\label{set}
\ea
Of these, the fourth, (2,3), corresponds to the dipole dependence of the form factors
$G_E$ and $G_M$ on $Q^2$, in which case two of the quarks reverses a spin upon the
proton non-spin-flip transition (the first number in parentheses is two); at the same
time, the proton spin-flip is due to the spin-flip for all three quarks (the second number
in parentheses is equal to three).

We obtain $G_E/G_M \sim 1$ for the (0,1) and (2,3) sets in (\ref{set}),
$Q^2 G_E/G_M \sim 4\, m^2$ for the (0,3) set, and $Q^2 G_M/G_E \sim 4 \,m^2$ for the (2,1) set.

\subsection{The set (0,1), $\bs{G_E}/\bs{G_M} \sim \bs{1}$, but both $\bs{G_E}$
and $\bs{G_M}$ behavior deviate from the dipole}

Let us consider the  first (0,1) set, corresponds to a proton non-spin-flip transition
$J_{p}^{\delta, \delta }$ for the case where there is no spin-flip for any of the three quarks
and corresponds to the proton transition $J_{p}^{-\delta, \delta }$ where spin-flip occurs
only for one quark. For this purpose we will make use of the above expressions
(\ref{Qprot}), (\ref{Qquark}) and (\ref{pQCD}).
It is convenient to represent this conceptual framework in the form of the following diagrams:
\ba
&& + \; \to \to *\to  \to \to \to \; + \nn \\
\label {Jdp}
 J_p^{\delta ,\delta }&=& - \; \to \to \to * \to \to \to \; - \quad \mbox{non-spin-flip}, \\
&& + \; \to \to \to  \to* \to \to \; +  \nn
\ea
\ba
&& + \; \to \to* \to  \to \to \to \; - \nn \\
\label {Jdm}
J_p^{-\delta ,\delta }&=& - \; \to \to \to*  \to \to \to \; -  \quad \mbox{spin-flip} \,.~~~\\
&& + \; \to \to \to  \to* \to \to \; +  \nn
\ea
The diagram in Eq. (\ref{Jdp}) corresponds to a proton non-spin-flip transition for
the case where there is no spin flip for any of the three quarks. It follows that, in this
case, the matrix element of the proton current must be proportional to $G_{E}$
[see Eq. (\ref{Qprot})]. As a result, we have
\ba
&&J_p^{\delta ,\delta }\sim G_{E}\sim 1 \times 1 \times 1\, \times \frac{1}{Q^6}\,,
\label{Ge0}
\ea
where the factors of unity correspond to non-spin-flip transitions [see Eq. (\ref{Qquark})]
for three pointlike quarks and $Q^6$ arises in the denominator owing to two
gluon and two quark propagators [see Eq. (\ref{pQCD})]. From here, we obtain
\ba
G_E \sim \frac{1}{Q^6}\;.
\label{Ge1}
\ea
The diagram in Eq. (\ref{Jdm}) corresponds to the transition where spin-flip occurs
for the up quark but does not occur for the two down quarks; in summary, this corresponds
to the proton spin-flip transition. According to Eqs. (\ref{Qprot}), the matrix element
of the proton current must be proportional to $\sqrt{\tau}\, G_{M}$ in this case. As a result, we have
\ba
\label{Gm0}
&&J_p^{-\delta ,\delta }\sim \sqrt{\tau} \,G_{M} \,\sim \sqrt{\tau} \times 1 \times 1 \times \frac{1}{Q^6}\,,
\ea
whence we obtain
\ba
G_{M} \sim   \frac{1}{Q^6}\; .
\label{Gmd}
\ea
The factor $\sqrt{\tau}$ on the right-hand side of Eq. (\ref{Gm0}) corresponds to the spin-flip
transition for the up quark [see Eq. (\ref{tau tau0})], while the two factors of $1$ correspond
to the non-spin-flip transition for the down quarks; two gluon and two quark propagators yield $Q^6$
in the denominators on the right-hand sides of (\ref{Gm0}) and (\ref{Gmd}).
As a result, we have
\ba
G_E \sim  \frac{1}{Q^6}, \, G_M \sim \frac{1}{Q^6}\,, \, \frac{G_E}{G_M} \sim1\, .
\label{set01}
\ea
Therefore, the form factor ratio $G_E/G_{M}$ behaves in just the same way
as in the dipole model. However, the dependence $G_E\sim 1/(Q^6)$ and the dependence
$G_M \sim 1/(Q^6)$ are not of the dipole character (the dipole dependence correspond to
$G_E\sim 1/Q^4$ and $G_M \sim 1/Q^4$).


\subsection{The set (0,3), dependence $\bs{G_E}/\bs{G_M} \sim \bs{4}\bs{m^2}/\bs{Q^2}$}

Let us consider the (0,3) set, in which case spin-flip transitions
for all three quarks contribute to $J_{p}^{-\delta, \delta }$.
For this purpose we write equalities similar to (\ref{Ge0}) and (\ref{Gm0}); that is,
\ba
&&J_{p}^{\delta ,\delta }\sim G_{E} \sim 1 \times 1 \times 1\, \times \frac{1}{Q^6}\;,\\
&&J_{p}^{-\delta ,\delta } \sim  \sqrt{\tau}\;G_{M}\sim \sqrt{\tau} \times\sqrt{\tau}
\times\sqrt{\tau} \, \times \frac{1}{Q^6}\;.
\ea
From here, we obtain
\ba
&& G_E \sim \frac{1}{Q^6}\;,\;\; G_M \sim \frac{\tau}{Q^6}\;,
\;\frac{G_E}{ G_{M}} \sim\frac{1}{\tau} \sim  \frac{4m^2}{Q^2}\;,\\
&& ~~~~ Q^2\, \frac{G_E}{ G_{M}} \sim 4m^2=\const
\label{nasBrodsky}
\ea
It follows that, for $Q^2>4 m^2$, the ratio $G_E/G_M$ becomes smaller than unity.
This is one possible way of violation of the dipole dependence in question. It is due to the
occurrence of the spin-flip process for all three quarks. At the same time, the dependence
that we obtained differs from the dependence (\ref{linfit}) observed at JLab.

Note that  the relation (\ref{nasBrodsky}) is sometimes called in the literature as the Brodsky saturation
law. Obviously really it correspond to the maximal possible number of the quarks spin-flip transition
in which case of the proton transition with spin-flip.

\subsection{The set (2,1), dependence $\bs{G_E}/\bs{G_M} \sim \bs{Q^2}/\bs{4m^2}$}

Let us now consider the (2,1) spin combination in the set (\ref{set}). It is generated
by spin-flip transitions for two quarks in the case of the contribution to
$J_p^{\delta ,\delta }$ and by spin-flip transitions for only one quark in the case of
the contribution to $J_p^{-\delta ,\delta }$. Following the same line of
reasoning as above, one can readily show that, for the (2,1) set, $G_E$ and $G_M$ have the form
\ba
&&G_{E} \sim  \frac{\tau}{Q^6} \sim \frac{1}{Q^4}\;, \; G_{M} \sim \frac{1}{Q^6}\; , \;
\ea
that is, the ratio $G_E / G_{M}$ behaves as
\ba
\frac{G_E}{G_{M}} \sim \tau \sim  \frac{Q^2}{4m^2}\;, \; Q^2\;\frac{G_M}{ G_{E}} \sim 4m^2=const\,.
\ea

\subsection{The set (2,3), $\bs{G_E}/\bs{G_M} \sim \bs{1}$,
dipole dependence of the form factors $\bs{G_E}$ and $\bs{G_M}$ on $\bs{Q^2}$}


Let us consider the (2,3) spin combination in the set (\ref{set}). It is generated by
spin-flip transitions for two quarks in the case of the contribution to
$J_p^{\delta ,\delta }$ and by spin-flip transitions for all three quarks in the case of
the contribution to $J_p^{-\delta ,\delta }$. In the case being considered, we have
\ba
&&J_{p}^{\delta ,\delta } \sim G_{E}\sim  \sqrt{\tau} \times \sqrt{\tau} \times 1 \, \times \frac{1}{Q^6}\;,\\
&&J_{p}^{-\delta ,\delta } \sim \sqrt{\tau} \,G_{M} \sim  \sqrt{\tau} \times  \sqrt{\tau}
\times  \sqrt{\tau}\,\times \frac{1}{Q^6}\;.
\ea
Whence we obtain
\ba
&& G_E \sim \frac{\tau}{Q^6}, \;\; G_M \sim \frac{\tau}{Q^6}\;, \\
&& G_E \sim \frac{1}{Q^4}, \; G_M \sim \frac{1}{Q^4}, \; \frac{G_E}{ G_{M}} \sim  1\;.
\ea
Thus, the dipole dependence in the behavior of the form factors arises owing to
the contribution by spin-flip transitions for two quarks in the case of the contribution to
$J_p^{\delta ,\delta }$ and by spin-flip transitions for all three quarks in the case of
the contribution to $J_p^{-\delta ,\delta }$. The dipole dependence can be realized
at high $Q^2$ in the case when the quark spin-flip transitions become dominant.
In other words it take place for the case when the number of quark transitions with spin-flip
is maximal, i.e. the saturation take place.

\subsection{Spin Parametrization for $\bs{G_E}/\bs{G_{M}}$}

The non-spin-flip and spin-flip proton-current amplitudes ($J_p^{\delta ,\delta }$
and $J_p^{-\delta ,\delta }$, respectively) can be represented as the linear combinations
\ba
\label{linge}
J_p^{\delta ,\delta }&=&\alpha_0\, J_q^{\delta,\delta} J_q^{-\delta,-\delta}  J_q^{\delta,\delta}
 + \alpha_2 \, J_q^{-\delta,\delta} J_q^{\delta,-\delta}J_q^{\delta,\delta},~~~~~~ \\
J_p^{-\delta ,\delta }&=&\beta_1 J_q^{-\delta,\delta} J_q^{\delta,\delta} J_q^{-\delta,-\delta}
 + \beta_3\, J_q^{-\delta,\delta} J_q^{\delta,-\delta} J_q^{-\delta,\delta},
 \label{lingm}
\ea
where the coefficients $\alpha_0$, $\alpha_2$, $\beta_1$ and $\beta_3$ have a clear
physical meaning and their indices determine the number of quarks undergoing spin-flip transitions and
contributing to proton non-spin-flip and spin-flip transitions. With the aid of Eqs. (\ref{linge})
and (\ref{lingm}), one can readily obtain a general expression for the ratio $G_E/G_{M}$.
The result is
\ba
&&\frac{G_E}{G_M}=
\frac{\alpha_0\, + \alpha_2\, \tau}{\beta_1 \, + \beta_3 \, \tau} \;.
\label{genform}
\ea
This expression may serve as a basis for constructing a spin parametrization and fits
to experimental data obtained by measuring the ratio $G_E/G_{M}$.

Because of the requirement that for the set (0,1) when quarks non-spin-flip transitions are dominant
the ratio $G_E/G_{M}\sim1$ hold at small $\tau$, the coefficients $\alpha_0$ and $\beta_1$
in Eq. (\ref{genform}) must obviously be close to unity: $\alpha_0 \sim 1$ and $\beta_1 \sim 1$.
With allowance for this comment, we expand the right-hand side of (\ref{genform}) in a power
series for $\tau$. As a result, we arrive at the law of a linear decrease in the ratio $G_E/G_{M}$
as $Q^2$ increases; this law agrees with (\ref{linfit}) established experimentally
in \cite{Gayou02}:
\ba
&&\frac{G_E}{G_M} \sim 1- \frac{(\beta_3-\alpha_2)}{4m^2} \; Q^2\;.
\label{linform}
\ea
Thus, the measurement of the ratio $G_E/G_M$ provides valuable insights into the proton
and to determine the number of its quarks whose spins were reversed.

\begin{figure}[ht!]
\hspace{-0.8cm}
\includegraphics[width=.47\textwidth]{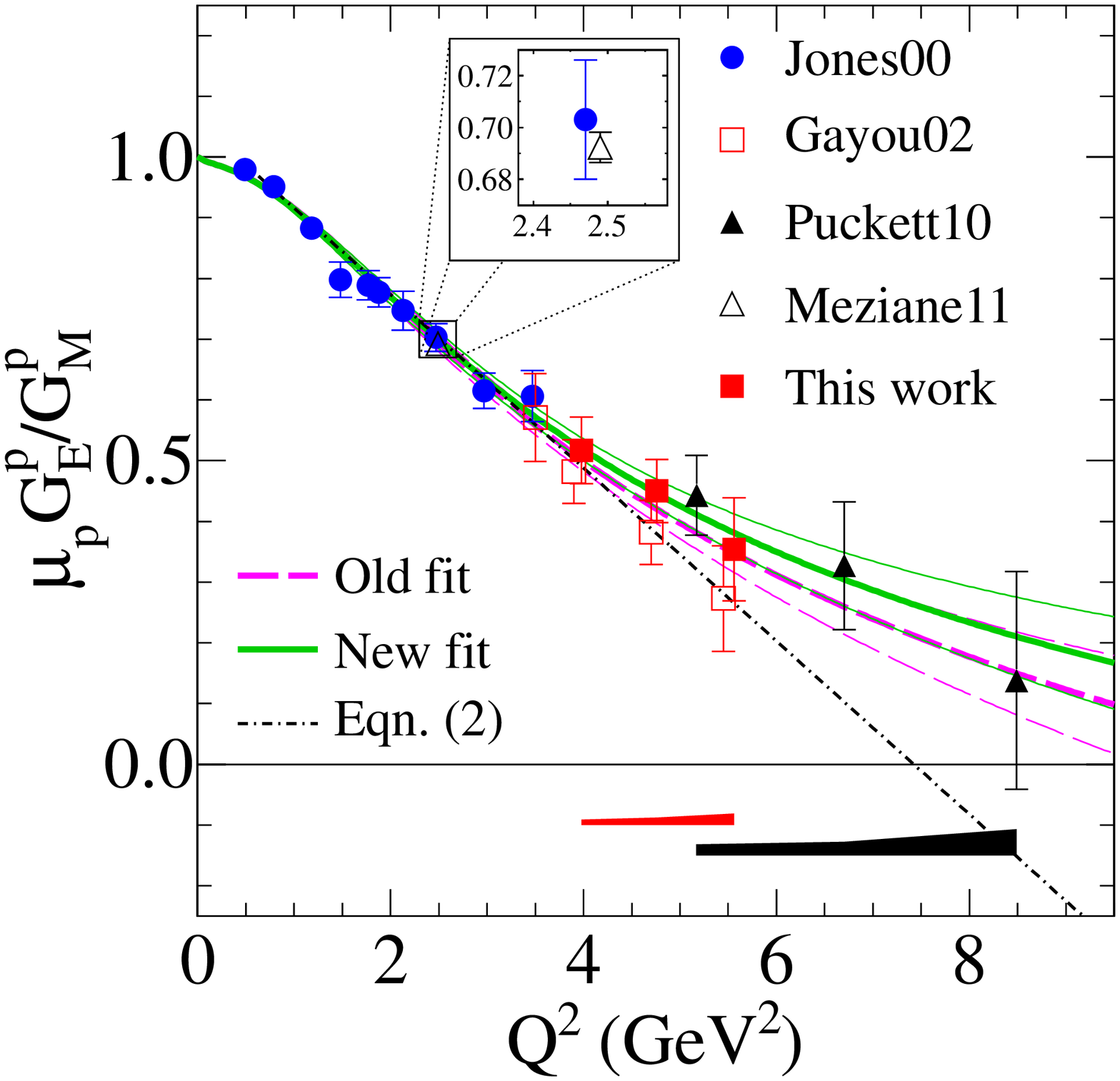}
\caption{ Polarization transfer data 
for $G_E^p/G_M^p$ from \cite{Jones00} (Jones00), \cite{Gayou02} (Gayou02), \cite{Puckett10}
    (Puckett10), \cite{Meziane11} (Meziane11) and work \cite{Puckett12} (Puckett12) (red squares).
Curves are global proton form factor fits using the originally published GEp-II data
\cite{Gayou02} (Old fit) and work \cite{Puckett12} (New fit). Both fits include
the GEp-III data. The linear fit of equation \eqref{linfit} is shown for comparison.}
\label{ratiofig}
\end{figure}

\section*{Summary and Conclusion}

The questions of how a dipole character of the dependence of the form factors $G_E$ and $G_M$ on the square of the
momentum transfer to a proton, $Q^2$, arise and why a violation of this dependence occurs, which was first
observed in a JLab polarization experiment, are investigated.
The answers to these questions could be obtained owing to the use of the simplest QCD concepts
of the proton structure and the results obtained by calculating the matrix elements of the proton
current in the case of non-spin-flip and spin-flip transitions for protons in
the DSB.
In the DSB, the form factors $G_E$ and $G_M$ are determined by the matrix elements $J^{\delta ,\delta }_{p}$,
and $J^{-\delta ,\delta }_{p}$ of the proton current in the cases of non-spin-flip and spin-flip
transitions for protons.
In an arbitrary reference frame, the relations between these matrix elements and the form factors
are $J^{\delta ,\delta }_{p}  \sim G_E$ and $J^{-\delta ,\delta }_{p} \sim  \sqrt{\tau}\, G_M$.
%
In considering the problem in question at the quark level, to obtain the leading contribution
to the matrix element of the proton current at $Q^2\gg 1$ $GeV^2$ we used the assumption that the respective
matrix element of the proton current is the product of three pointlike-quark-current amplitudes
(having the form (\ref{Qquark})), of two hard gluon propagators ($\sim 1/Q^2$) and of two hard quark
propagators ($\sim 1/Q$), see Eq. (\ref{pQCD}).

Using relations (\ref{Qprot}) and (\ref{Qquark}) and (\ref{pQCD}) we obtained the law
of the linear decrease in the ratio $G_E/G_{M}$ as $Q^2$ increases (\ref{linform}) 
established experimentally in \cite{Gayou02}.
Besides we had considered all the combinations in the set (\ref{set}) and find corresponding
dependence for $G_E$ and $G_M$ on $Q^2$.

At $\tau \ll 1$ ($\tau \gg1$) the quark transition without (with) spin-flip dominate (see Eqs. (\ref{Qquark}))
the set (0,1) with the minimal number of spin-flip quarks (here $G_E/G_{M} \sim 1$
but $G_E, G_M \sim 1/Q^6$) must occur at $\tau \ll 1$.
We have shown that the dipole dependence of the form factors $G_E$ and $G_M$ on $Q^2$ is realized
in the set (2,3) when the number of spin-flipped quarks is maximal, i.e. when $\tau \gg 1$.
In the case when $\tau < 1$ the ratio $G_E/G_{M}$ decreases linearly with increasing of $Q^2$.
This linear dependence is caused by the contributions to $J^{\delta ,\delta }_{p}$ from spin-flip
transitions for two quarks or by the contribution to $J^{-\delta ,\delta }_{p}$ from spin-flip
transitions for all three quarks constituting the proton but the fraction of such contributions
at $\tau < 1$ must be relatively small, see Eq. (\ref{genform}) and Eq. (\ref{linform}).

In Fig. (\ref{ratiofig}) (we take it from paper \cite{Puckett12}) all the results for the ratio
$G_E/G_{M}$ obtained in JLab experiments for the region $0.5 \leq Q^2 \leq 8.5 ~ \GeV^2$
are presented.
As can be seen from this figure, in the lower boundary of this region (near the $Q^2\approx 1 ~ \GeV^2 $)
the result $R\approx 1$ is in agreement with one followed from Eqs. (\ref{set01}) for the set (0,1).
Our results for $R$ in the region of $\tau < 1$ ($1 \leq Q^2 \leq 3.0 ~ \GeV^2$) as well
are consistent with experimental data.

We believe that the presented above interpretation can be considered as a possible solution
of the $G_E/G_{M}$ problem. One of our predictions is the realization (restoration) of a dipole
dependence of form factors and the value $R = 1$ for higher values of $Q^2$ (at $Q^2\gg 4 m^2$).

\vspace{0.0cm}

\section*{Acknowledgments}
We are deeply grateful to E.A. Tolkachev for helpful discussions and to V.L. Chernyak and V.S. Fadin for
critical remarks. This work was supported in part by the Belarusian Republican Foundation
for Fundamental Research,  grant F10D-005.

\appendix
\section{Calculation of QED matrix elements in the DSB}
\label{DSBmethod}
\subsection*{Preface}

In the DSB the little Lorentz group (the little Wigner group \cite{Wigner}) common
for the initial and final states, is being realized \cite{FIF70,GL}. This brings the spin operators
of {\em in-}\ and {\em out-}particles to coincidence and makes it possible to separate the interactions
with and without change in the spin states of the particles involved in the reaction
in the covariant form and, thus, to trace the dynamics of the spin interaction.
The spin states of massless particles in the DSB coincides up to a sign with the helicity basis
\cite{GS89,GS98}; in this case, the DSB formalism is equivalent to the CALKUL group method \cite{CALKUL}.
In contrast to methods of CALKUL-group etc, the developed approach is valid both for massive fermions
and for massless ones. There occur no problems with accounting for spin-flip
amplitudes in it. No auxiliary vectors are to be introduced in DSB. Just
4-momenta of particles participating in reaction are required in it to
construct the mathematical apparatus for amplitude calculation.

In the DSB, Wigner rotations, which are purely kinematical in nature, are separated from the
amplitudes. This leads to maximal simplification of the mathematical structure
of the matrix elements in the DSB, and the resulting expressions give the truest
reflection of the physical essential of spin phenomena.

In the used by us Bogush-Fedorov covariant approach \cite{GL} the calculation of matrix
elements of the form (\ref{MQED}) reduces to evaluating the trace:
\be
 M^{\pm\delta,\delta} = Tr (P_{21}^{\pm\delta,\delta} Q ) \; , \; P_{21}^ {\pm\delta,\delta} =
u^{\delta}(q_{1}) \; \overline {u}^{~\pm \delta}(q_{2}) \; . \\
\label{MQED3}
\ee
To construction of the operators $P_{21}^{\pm\delta,\delta}$ we need to know
\begin{itemize}
\item
the projection operators of the particle states:\\
$\tau_1^{\delta}=u^{\delta}(q_1)\,\overline{u}^{\delta}(q_1)$ and
$\tau_2^{\delta}=u^{\delta}(q_2)\,\overline{u}^{\delta}(q_2)$;

\item
the operator $T_{21}$ (and its inverse operator $T_{12}$, $T_{12}=T_{21}^{-1}$, $T_{21}T_{12}=1$) for the transition
from the initial to the final state without spin-flip:
$u^{\delta}(q_2)=T_{21}u^{\delta}(q_1)$, $u^{\delta}(q_1)=T_{12}u^{\delta}(q_2)$,
$\overline{u}^{\delta}(q_{2}) = \overline {u}^{\delta}(q_{1}) T_{12}$;

\item
the raising and lowering spin operators in the case of transitions with spin flip.
They given by Eq.(\ref{spop6b}).
\end{itemize}

\subsection{The projection operators of particles with spin 1/2 in the DSB}

Let us consider the projection operators of particles with spin 1/2,
$\tau^{\delta}_i = u^{\delta}(q_{i}) \; \overline {u}^{\delta}(q_{i})$ \cite{AB}:
\ba
\tau^{\delta}_i = 1/2 (\hat q_i + m) ( 1 - \delta \gamma^{5} \hat s_i ) \; ,
\label{tau}
\ea
where $q_i$ and $s_i$ are 4-momenta and spin 4-vectors with $q_i^2=m^2$ and $s_i^2=-1$,
$q_{i}s_i=0, i=(1, 2)$. In the DSB (\ref{DSB}) the operators $\tau^{\delta}_{i}$ (\ref{tau})
have the form \cite{GS89,GS98}:
\ba
\label{tau1}
\tau^{\delta}_{1}& = & 1/2\, [ m +  \xi_{+} \hat b_{0} - \xi_{-} \hat b_{3}+ \\
&+& \delta \gamma^{5} \,( \xi_{-} \hat b_{0} - \xi_{+} \hat b_{3} - m \hat b_{3} \hat b_{0} ) ] \nn \,,\\
\label{tau2}
\tau^{\delta}_{2} &=& 1/2 \, [ m +  \xi_{+} \hat b_{0} + \xi_{-} \hat b_{3} -\\
&-& \delta \gamma^{5}  ( \xi_{-} \hat b_{0} + \xi_{+} \hat b_{3} +  m \hat b_{3} \hat b_{0} ) ]  \nn\,.
\ea
Here 4-vectors $b_0, b_3$ and $q_+, q_-$ are defined by Eq. (\ref{OBV}),
$\xi_{\pm} = \sqrt{\pm q_{\pm}^2}/2$. Owing to (\ref{spop6b}),
the spin parts of the projection operators for particles 1 and 2 in the DSB can be made
identical, and so we have \cite{GS98}:
\ba
\tau^{\delta}_{i} = - 1/4 \; (\hat q_{i} + m ) \; \hat b_{\delta} \;
 \hat b_{\delta}^{\ast} \; ,
\label{coin}
\ea
where $b_{\delta}^{\ast}=b_{-\delta} = b_{1} - i \delta b_{2}$ and
$b_{\delta} b^{\ast}_{\delta}=-2$. Here 4-vectors $b_1, b_2$ are defined by Eq. (\ref{OBV}).

In the massless case the projection operators $ \tau^{\delta}_{1}$ and
$ \tau^{\delta}_{2}$ (\ref{tau1}) and (\ref{tau2}) take the form \cite{GS89,GS98}:
\ba
\tau^{\delta}_{1} = \hat q_{1} ( 1 - \delta \gamma^{5})/2 \; ,
\tau^{\delta}_{2} = \hat q_{2} ( 1 + \delta \gamma^{5})/2 \; .
\label{tau120}
\ea
It is easy to show that the operators $\tau^{\delta}_{1}$ and
$\tau^{\delta}_{2}$ (\ref{tau120}) satisfy the relations:
\ba
\label{ga5tau}
\gamma^{5} \tau^{\delta}_{1} &=& \delta \, \tau^{\delta}_{1} \; ,\;\;\;\;
\gamma^{5}\tau^{\delta}_{2}=-\delta \; \tau^{\delta}_{2} \; ,\\
\tau^{\delta}_{1} \; \gamma^{5} &=& - \delta \; \tau^{\delta}_{1}  \; ,\;
\tau^{\delta}_{2} \; \gamma^{5} = \delta \; \tau^{\delta}_{2} \; .
\label{tauga5}
\ea
Remembering that in the massless case, the matrix $\gamma^5$ is the helicity operator,
we come to the conclusion, that in the massless case in the DSB the initial state is a helicity state,
and the final state has negative helicity.

\subsection{The operator $\bs {T_{21}} $ for the transition from the initial to the final state without spin-flip}

The bispinors of the initial and final states of the particles,
$u^{\delta}(q_{1})$ and $u^{\delta}( q_{2})$, can be related to each other
by using the transition operators $T_{21}$ and $T_{12} = T_{21}^{-1}$ \cite{GS89,GS98}:
\ba
u^{\delta}( q_{2} ) = T_{21} \; u^{\delta}(q_{1}) \; , \overline {u}^{\delta}
 (q_{2}) = \overline {u}^{\delta}(q_{1}) \; T_{12} \; ,
\label{uut21}
\ea
which in the DSB have the form \cite{GS89,GS98}:
\ba
T_{21} = \frac{ 1 + \hat v_{2} \hat v_{1}}{\sqrt{ 2 (v_{1} v_{2} + 1)}} \; ,
\; T_{12} = \frac{ 1 + \hat v_{1} \hat v_{2}}{\sqrt{ 2 ( v_{1} v_{2} + 1)}} \; ,
\label{t21vi}
\ea
where $v_i=q_i/m$. Note that the Dirac equation can be used to reduce the transition
operators $T_{21}$ and $T_{12}$ (\ref{t21vi}) to the same form \cite{GS89,GS98}:
\ba
T_{21} = T_{12} = \hat b_{0} \; .
\label{t21b0}
\ea

\subsection{The construction
of operators $\bs{ P_{21}^{\pm\delta,\delta}=u^{\delta}(q_{1}) \overline {u}^{~\pm \delta}(q_2)}$
}

In the papers \cite{GS89,GS98} we have constructed the operators $P_{21}^{\pm\delta,\delta}
=u^{\delta} (q_{1}) \,\overline {u}^{~\pm \delta}(q_{2})$ (\ref{MQED2}) used to calculate the
DSB amplitudes in the case of transitions without and with spin-flip. They can be easy evaluated
by the next way:
\ba
&&~~P_{21}^{\delta,\delta} = u^{\delta}(q_{1})\,\overline {u}^{\delta}(q_{2})
= u^{\delta}(q_{1}) \,\overline {u}^{\delta}(q_{1}) T_{12} = \tau_{1}^{\delta}\, T_{12} ,\nn~\\
\label{p21pp}
&&P_{21}^{-\delta,\delta} = u^{\delta}(q_{1})\, \overline {u}^{~-\delta}(q_{2})
= \sigma^{+\delta} u^{-\delta}(q_{1})\, \overline {u}^{~-\delta}(q_{2}) =\nn \\
&& ~~~~~~~=\sigma^{+\delta} P_{21}^{~-\delta,-\delta} .
\label{p21pm}
\ea
The operators $P_{21}^{\pm\delta,\delta}$ (\ref{p21pm}) determine the structure of the
spin dependence of the matrix elements (\ref{MQED}) in the case of transitions
without spin-flip ($M^{\delta,\delta}$) and with spin-flip ($M^{-\delta,\delta}$).
Their explicit form in the DSB can easily be obtained by using Eqs. (\ref{spop6b}), (\ref{tau1}), (\ref{tau2}),
and (\ref{t21b0}):
\ba
\label{2.5}
P_{31}^{\delta,\delta} &=& [\, \xi_{+} + m \,\hat b_{0} - \xi_{-} \hat b_{3} \, \hat b_{0} + \\
&+&\delta \gamma^{5} ( \xi_{-} - m \,\hat b_{3} - \xi_{+}\hat b_{3} \,\hat b_{0} ) ]/2 ,\nn\\
P_{31}^{-\delta,\delta} &=& - \delta \, ( \, \xi_{-} + m \, \hat b_{3}
 + \xi_{+} \, \delta \, \gamma^{5} \, )\; \hat b_{\delta} /2\, .
\label{2.6}
\ea
Equations (\ref{2.5}) and (\ref{2.6}) can be used to calculate the matrix elements, both
with and without spin-flip, for arbitrary $Q$. In particular, if the
interaction operator reduces to the form
\ba
Q = \hat A_1 + \gamma^{5} \; \hat A_2 \; ,
\label{2.7}
\ea
where $A_1$ and $A_2$ are any 4-vectors, then for the matrix elements (\ref{MQED}) we will
have:
\ba
&&M^{\delta,\delta} = 2 m \, ( A_1 b_{0} + \delta \, A_2 b_{3} \, ) \; ,\\
\label{2.8}
&&M^{-\delta,\delta} = 2\, [-\delta \xi_{-} \, (A_1 b_{\delta}) + \xi_{+} \,( A_2 b_{\delta})] \, .
\label{2.9}
\ea
Equations (\ref{2.5}) and (\ref{2.6}) can be written more compactly by using the operators
(\ref{coin}) and (\ref{t21b0}), and also the expressions \cite{GS89,GS98}:
\ba
\hat b_{3} \hat b_{0} \hat b_{\delta} = - \delta \gamma^{5} \hat b_{\delta} ,
 \gamma^{5} \hat b_{\delta} \hat b_{0} = \delta \hat b_{3} \hat b_{\delta}  ,
\gamma^{5} \hat b_{\delta} \hat b_{3} = \delta \hat b_{0} \hat b_{\delta}  \,.~~~
\label{2.10}
\ea
As a result, for the operators $P_{21}^{\pm\delta,\delta}$ we have \cite{GS89,GS98}:
\ba
\label{2.11}
P_{21}^{\delta,\delta} &=& ( \hat q_{1} + m ) \; \hat b_{\delta} \;
 \hat b_{0} \; \hat b_{\delta}^{\ast} /4 \; ,\\
P_{21}^{-\delta,\delta} &=& \delta \; ( \hat q_{1} + m) \;
 \hat b_{\delta} \; \hat b_{3} /2\;  .
\label{2.12}
\ea
In the massless case ($q_{1}^{2}=q_{2}^{2}=0$) the operators $P_{21}^{\pm
 \delta,\delta}$ in (\ref{2.5}) and (\ref{2.6}) take the form \cite{GS89,GS98}:
\ba
&& P_{21}^{\delta,\delta} =  \xi ( 1 + \delta \gamma^{5} ) ( 1 + \hat b_{0} \hat b_{3} )/2 \; ,\\
&& P_{31}^{-\delta,\delta} = - \delta \xi ( 1 + \delta  \gamma^{5} )\, \hat b_{\delta}/2 \; ,
\label{2.17}
\ea
where $ \xi = \xi_{+} = \xi_{-} = \sqrt{ q_{1} q_{2} /2}$.

\section{Standard and alternative methods for calculation $\bs{ep \to ep}$ process cross sections}
\label{sec:standard method}

The cross section (\ref{Ros}) 
can be represented as the sum of the cross sections without spin-flip
($\sigma^{\delta,\delta}$) and with spin-flip ($-\sigma^{\delta,\delta}$) of the initial
proton:
\ba
\label{Ros1}
&&\frac{d\sigma} {d\Omega}= \kappa \left(\,G_E^2 +\frac{\tau}{\varepsilon}\,G_M ^2\right)\,
=\kappa \,(\sigma^{\delta,\delta} + \sigma^{-\delta,\delta}),\;\;\\
&&~~~~~~~\sigma^{\delta,\delta}=G^2_E ,\;\; \sigma^{-\delta,\delta}=\frac{\tau}{\varepsilon} \, G^2_M\;.
\label{Ros2}
\label{quadraty}
\ea
where $\kappa$ is the factor in front of the parentheses in Eq. (\ref{Ros}).
At the same time the axes of the spin projections $\vecc c_{1}$ and $\vecc c_{2}$ should be coincide with
the direction of final-proton motion: $\vecc c_{1} =\vecc c_{2}=\vecc n_2$ and the spin
4-vectors $s_{1}$ and $s_{2}$ for initial and final protons must have the form
\ba
s_1=(0,\vecc n_2 ), \, s_2= (|\vecc v_2|, v_{20}\, \vecc {n_2})\,.  
\ea
The terms $\sigma^{\delta,\delta}$ and $\sigma^{-\delta,\delta}$ in Eq. (\ref{Ros1}), (\ref{Ros2})
are the cross sections without and with the spin-flip for the case
when the initial and final protons are fully polarized in the direction of the motion of the final proton.
For the case when $\vec c_{1}=\vec n_2$ and $\vec c_{2}=\vec n_2$ we have $\sigma^{\delta,\delta}$
and for the case when $\vec c_{1}=\vec n_2$ and $\vec c_{2}=-\vec n_2$ we have $\sigma^{-\delta,\delta}$.

Let us remind that the general form for spin 4-vectors $s_{1}$ and $s_{2}$ for protons with 4-momentum
$q_1$, $q_2$ is:
\ba
s_i=(s_{0i}, \vecc s_i), s_{0i}=\vecc v_i\, \vecc c_i,  \vecc s_i =\vecc c_i
+ \frac{(\vecc c_i \vecc v_i)\,\vecc v_i}{1+v_{0i}},
\ea
where $v_i=(v_{0i}, \vecc v_i)=q_i/m, i=1, 2$.

To prove the relation (\ref{Ros1}), (\ref{Ros2}) there are two additional ways:
\begin{itemize}
\item
Using the standard method calculation for QED processes cross sections \cite{AB}.
\item
With help of book \cite{HM} by F. Halzen and A. Martin "Quarks and leptons. An Introductory Course
in Modern Particle Physics", EXERCISE 8.7, Page 178, 1984 (in English); Page 214, 1987 (in Russian).
\end{itemize}

\subsection{Standard method for calculation $\bs{ep \to ep}$ process cross sections}

Evaluation of the cross section for the process $ep \to ep$  reduces to the calculation
of the square modulus of the matrix element (\ref{Mepep}) for this process:
\ba
\label{sechenie}
&&\sigma \sim |M_{ep\to ep}|^2 = |\overline{u}(p_{2}) \gamma^{\mu} u(p_{1}) \cdot \overline{u}(q_2)
\Gamma_{\mu}(q^{2}) u(q_1) |^2\, .\nn
\ea
In the standard method \cite{AB} this calculation of $\sigma$ with taken into account the polarization
of initial in final protons reduces to determination of product of lepton ($L^{\mu \nu}$) and
proton ($W_{\mu \nu}$) tensors
\ba
\label{sigma}
&&\sigma_{s_1,s_2} \sim L^{\mu \nu} W_{\mu \nu}, \\
\label{Lmunu}
&&L^{\mu \nu}=2 \cdot \Tr(\tau_2^e \gamma^\mu \tau_1^e \gamma^\nu),\\
\label{Wmunu}
&& W_{\mu \nu}=\Tr(\tau_2^p \,\Gamma_\mu \tau_1^p \,\overline{\Gamma}_\nu)\,,~~~
\ea
with
\ba
&& \tau_1^e=\frac{1}{2}(\hat{ p_1} +m_e),\, \tau_2^e=\frac{1}{2}(\hat{ p_2} +m_e)\,,\nn\\
&& \tau_1^p=\frac{1}{2}(\hat{ q_1} +m)(1-\delta_1\gamma_5 \hat{s}_1)\,,\nn \\
&& \tau_2^p=\frac{1}{2}(\hat{ q_2} +m)(1-\delta_2\gamma_5 \hat{s}_2)\,.\nn
\ea
Lepton tensor $L^{\mu \nu}$ (\ref{Lmunu}) have the form
\ba
\label{Lmunu1}
L^{\mu \nu}=2\,(p_1^{\mu} p_2^{\nu} + p_2^{\mu} p_1^{\nu})+q^2 g^{\mu \nu}\,. 
\ea
Tensor $L^{\mu\nu}$ in terms $p_+=p_2+p_1; p_-=p_2-p_1$ have a form
\ba
\label{Lmunu2}
L^{\mu \nu}_2 \equiv L^{\mu \nu}=p_+^{\mu} p_+^{\nu} - p_-^{\mu} p_-^{\nu}+q^2 g^{\mu \nu}\,.
\ea
In this equation the term $p_-^{\mu} p_-^{\nu}$ can be safely omitted as far as it do not contribute
to the cross section of process (\ref{sigma}). It is the consequence of the gauge invariance of QED amplitudes.
As a result for the lepton tensor we obtain a new, compact expression
\ba
\label{Lmunuc}
L^{\mu \nu}_c \equiv L^{\mu \nu}=p_+^{\mu} p_+^{\nu} + q^2 g^{\mu \nu}\,.
\ea
Using the representation (\ref{Gamu2}) for $\Gamma_{\mu}(q^2)$ and the definition of Dirac
formfactor in terms of the Sachs ones
\ba
F_1=\frac{G_E+\tau G_M}{1+\tau}=\frac{4m^2}{q_+^2}\,(G_E+\tau G_M)\,,
\ea
we obtain for tensor $W_{\mu\nu}$
\ba
\label{Wpmunu}
W_{\mu \nu} \equiv W_{\mu \nu}^{\delta_1 \delta_2}=\frac{1+\delta_1 \delta_2}{2}\,W_{\mu \nu}^{\delta, \delta}
+ \frac{1-\delta_1 \delta_2}{2}\,W_{\mu \nu}^{-\delta, \delta},~~~
\ea
with
\ba
\label{wpp1}
&&W_{\mu \nu}^{\delta, \delta}=\frac{4m^2G_E^2}{q_+^2}\,(q_+)_{\mu} (q_+)_{\nu}\,,\\
\label{wpm1}
&&W_{\mu \nu}^{-\delta, \delta}=\frac{4m^2\tau G_M^2}{q_+^2}\,
 \{ (q_+)_{\mu} (q_+)_{\nu} -q_+^2 g_{\mu \nu} + \\
&& + (q_-)_{\mu} (q_-)_{\nu}\,q_+^2/q_-^2
-4i\delta \varepsilon_{\mu \nu \rho\sigma}q_-^{\rho}q_+^{\sigma}\sqrt{q_+^2}/\sqrt{-q_-^2}
\}\,, \nn
\ea
where we as well can omit the term $(q_-)_{\mu} (q_-)_{\nu}$.

Note that for the case of unpolarized leptons (initial and the scattered) the asymmetry part
of the tensor $W_{\mu \nu}^{-\delta, \delta}$ (or the imaginary part of it)
in (\ref{wpm1}) as well do not contribute to the cross
section of process $e p\to e p$. So for tensors $W_{\mu \nu}^{\delta, \delta}$ and
$W_{\mu \nu}^{-\delta, \delta}$, which corresponds
to the cases with spin-flip an without spin-flip, for the unpolarized leptons we have
\ba
\label{wpp2}
&&W_{\mu \nu}^{\delta, \delta}=\frac{4m^2G_E^2}{q_+^2}\,(q_+)_{\mu} (q_+)_{\nu}\,,\\
\label{wpm2}
&&W_{\mu \nu}^{-\delta, \delta}=\frac{4m^2\tau G_M^2}{q_+^2}\,
\{ (q_+)_{\mu} (q_+)_{\nu} -q_+^2 g_{\mu \nu}\}\,. ~~
\ea
Forming the product of leptonic tensor (\ref{Lmunuc}) and the proton one (\ref{Wpmunu}) with
(\ref{wpp2}), (\ref{wpm2})) we obtain:
\ba
\label{Wppdd}
&&\sigma_{s_1,s_2} =\frac{(1+\delta_1\delta_2)}{2} W^{\delta,\delta}_{e p \to e p}
+\frac{(1-\delta_1\delta_2)}{2}\,W^{-\delta,\delta}_{e p \to e p},~~~~~~~~\\
\label{Wep pp2}
&&W^{\delta,\delta}_{ep\to ep}= \frac{4m^2G_E^2}{q_+^2} \,[(p_+q_+)^2+q_+^2q_-^2]\,,\\
&&W^{-\delta,\delta }_{ep \to ep}=\frac{4m^2\tau G_M^2}{q_+^2} [(p_+q_+)^2 - q_+^2 (q_-^2+4m_e^2)]\, .
\label{Wep pm2}
\ea
Thus, the differential cross section for the $ep \to ep $ process
naturally splits into the sum of two terms containing only the squares
of the Sachs form factors and corresponding to the contribution of transition
without ($\sim G_{E}^2$) and with ($\sim G_{M}^2$) proton spin-flip.

With the help of the matrix elements of the proton current (\ref{Jepep-pp}), (\ref{Jepep-pm}) calculation
probability of the process $ep \to ep$ can be reduced to calculation of the trivial trace:
\ba
\label{Tsquare}
&&\mid T \mid^2=\frac{4m^2}{q^4}\, \frac{1}{8}\,\sum_{\delta} Tr (G^2_E(\hat p_2+m_e)
\hat b_0(\hat p_1 +m_e)\hat b_0+ \nn \\
&&~~~~~~~~~~~~~+ \tau \,G^2_M(\hat p_2 +m_e)\hat b_{\delta}(\hat p_1 +m_e)\hat b^*_{\delta})\;.\nn
\ea
The expression for $|T|^2$ leads to the cross section, which coincides with
result in \cite{AB}:
\ba
\label{6.1}
&& ~~~~~~d \sigma = \frac{ \alpha^{2} d o}{4w^2} \frac{1}{1+\tau} \, ( \, G_{E}^{2} \, Y_{I}
+ \tau \; G_{M}^{2} \, Y_{II} \, )\, \frac{1}{q^{4}}\, , \\
&& Y_I=(p_+ q_+)^2+q_+^2q_-^2, \; Y_{II}=(p_+ q_+)^2-q_+^2(q_-^2+4 m_e^2)\;.   \nn
\ea

In our paper \cite{GKB2008} based on the use of the expression (\ref{Wppdd}) a new method
of measuring of the Sachs form factors was suggested. It was shown that they can be determined
separately and independently by direct measurements of the cross sections without
and with spin-flip of the initial proton, which should be at rest
and fully polarized in the direction of the motion of the scattered proton.

Using the matrix elements of the proton current in DSB (\ref{Jepep-pp}), (\ref{Jepep-pm})
for the proton tensor $W_{\mu \nu}^{\delta_1, \delta_2} $ one can construct an another
equivalent and compact expression:
\ba
\label{WmunuDSB}
&&W_{\mu \nu}^{\delta_1, \delta_2}=4m^2\left [\frac{(1 + \delta_1 \delta_2)}{2}\, G_E^2 (b_0)_{\mu} (b_0)_{\nu}+ \right. \nn \\
&&~~~~~\left. +\, \frac{(1 - \delta_1 \delta_2)}{2}\, \tau_p \,G_M^2 (b_{\delta})_{\mu} (b^{\ast}_{\delta})_{\nu}\right ]\,.
\ea

\subsection{
An alternative method of calculation of the spin-flip and non-spin-flip proton current matrix elements
}
\begin{figure}[h!]
\vspace{0.0cm}
\hspace{-1.50cm}
\includegraphics[width=10.0cm]{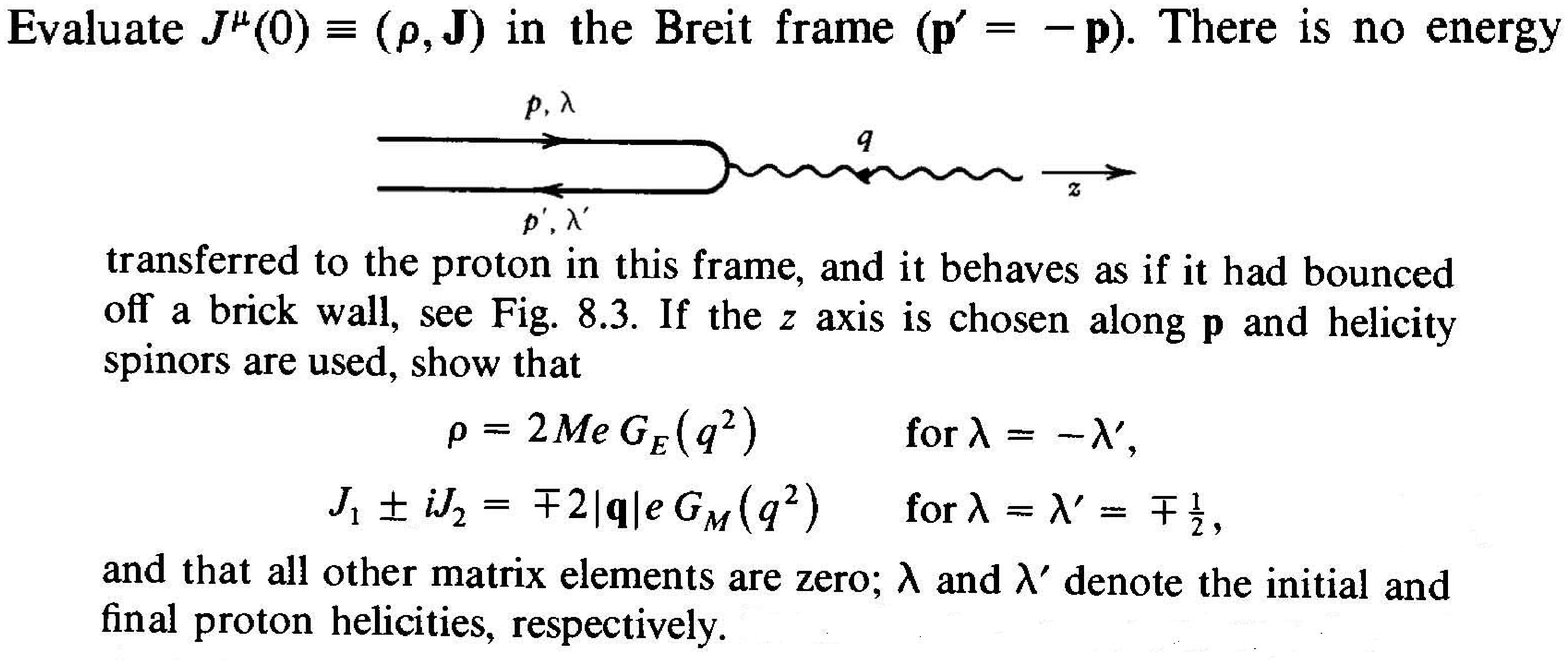}
\vspace{-0.50cm}
\caption
{\bf
Exercise 8.7 at page 178 from book of F. Halzen and A. Martin \protect\cite{HM}. 
}
\label{epgfig1b}
\end{figure}

To prove the correctness of the results obtained in the DSB for the proton current matrix elements
(\ref{Jepep-pp}), (\ref{Jepep-pm}) we propose to consider here
Exercise 8.7 at page 178 from book of F. Halzen and A. Martin \cite{HM}  (Fig. 8.3 also extracted
from this book and show at Figure \ref{epgfig1b}).
In this exercise one suggests to consider the matrix elements of the proton
current in the Breit reference frame and show that the proton transition with helicity-flip
(without helicity-flip) are determined by only the Sachs electric formfactor $G_E$
(magnetic form factor $G_M$).

From this picture, we see that in the Breit-system a transition with (without) a change in the sign
of helicity is the transition without (with) spin-flip of the proton
\ba
&& J^{-\lambda,\lambda}_{\mu}=J^{\delta,\delta }_{\mu} = 2\, e\, M \,G_{E} \,(b_{0})_{\mu} \, , \, \\
&& J^{\lambda,\lambda}_{\mu}=J^{-\delta,\delta }_{\mu}=- 2 e\, \delta |\vecc q| \,G_{M}\,
(b_{\delta })_{\mu}\, , \\
&& |\vecc q|=\sqrt{Q^2}, \nn
\ea
where
\ba
\label{BAA}
b_{0}=(1,0,0,0), b_1=(0,1,0,0), b_2=(0,0,1,0), \\
b_3=(0,0,0,1), b_{\delta} = b_{1} + i \delta b_{2}, \delta=\pm 1\,. \nn
\ea
\vspace{0.20cm}
Below we will dropped the factor $e$ in matrix elements and denote by the letter $m$ of the proton mass:
\ba
 J^{\delta,\delta }_{\mu} = 2\, m G_{E} \,(b_{0})_{\mu}  ,
 J^{-\delta,\delta }_{\mu}=- 2\,m \delta \sqrt{\tau} G_{M}\,(b_{\delta })_{\mu} .~~~
\label {FFHM}
\ea
In the Breit system where $q_1=(q_0, -\vecc q), q_2=(q_0, \vecc q)$ and the spin states of
the initial and final protons are helicity, so they spin four-vectors  $s_{1}$ and $s_{2}$ have the form:
\ba
s_1=(-|\vecc v|, v_0 \vecc n_2), \,
s_2= (|\vecc v|, v_{0} \vecc n_2)\, , \vecc n_2=  \vecc {q_2}/|\vecc q_2|\,.\nn
\ea

For the transition from the Breit system to an arbitrary reference frame we need
to make the Lorentz transformation. Instead for this purpose we will construct 4-vectors $b_A$
(\ref{BAA}) through the 4-momenta of participating in the reaction particles. The unit 4-vectors
$b_0$ and $b_3$ can be written as the normalized per unit the sum and difference between
the momenta of final and initial protons:
\ba
\label{b0}
&&b_0=\frac{q_+}{\sqrt{q_+^2}}, \,b_3=\frac{q_-}{\sqrt{-q_-^2}}\;,\\
\label{bd}
&& (b_1)_{ \mu} = \varepsilon_{\mu \nu \kappa \sigma}b_0^{\nu}b_3^{\kappa}b_2^{\sigma},
(b_{2})_{\mu} = \varepsilon_{\mu \nu \kappa \sigma}b_0^{\nu}b_3^{\kappa}p_1^{\sigma}/\rho\,,~~~~\\
&&q_+=q_1+q_2=(2q_0, 0,0,0),  \Rightarrow b_0=(1, 0,0,0),~~~\nn\\
&& q_-=q_2-q_1=(0,0,0,2q), \Rightarrow b_3=(0,0,0,1), \nn
\ea
The matrix elements of the proton current (\ref{FFHM}) by using (\ref{b0}), (\ref{bd}) coincide with
results (\ref{Jepep-pp}), (\ref{Jepep-pm}) in DSB and are valid in arbitrary reference frame.

\section{Virtual-photon polarization in the reaction $\bs{ep \to ep} $}
\label{Linpol}

For the study of the virtual photon polarization in the process
$ep \to ep$ usually used leptonic tensor \cite{Rekalo68,Dombey} which defined as
\ba
L_{\mu \nu}=J_\mu J_\nu^*, \; J_\mu=\overline{u}(p_2)\gamma_\mu u(p_1) \, .
\ea
The interpretation of the results is considerably simplified if the tensor $L_{\mu \nu}$
is expressed in terms of the longitudinal and transverse polarization vectors of the virtual photon.
The relevant expressions can be found in the literature. It should be noted however two disadvantages
of such expressions: 1) they disregard mass of the electron, which is of course justified
at ultrarelativistic electron energies and large squared 4-momentum of the virtual photon;
2) they have a noncovariant form. A leptonic tensor that is free from the above flaws
was constructed in \cite{GLev}. In this case, was used the explicit form for the matrix
elements of the electron current in the DSB.

From Eqs. (\ref{Jepep-pp}), (\ref{Jepep-pm}), we can write
the matrix elements of the electron current in DSB \cite{Sik84,GS98}
\ba
\label {Jel-pp}
&&(J^{\delta_e,\delta_e }_{e} )_{\mu} = 2 m_e \,( a_{0} )_{\mu} \, , \\
\label {Jel-pm}
&&( J^{-\delta_e,\delta_e }_{e} )_{\mu}=- 2  m_e \,\delta_e \sqrt{\tau_e} \,(a_{\delta_e })_{\mu}\, ,
\label {FFSep-c}
\ea
with orthonormal basis of 4-vectors $a_{A} \,( A=0, 1, 2, 3)$, constructed from 4-momenta of the electrons:
\ba
\label{OBVe}
&& ~~~~a_0=p_+/\sqrt{p_+^2}, \; a_{3} = p_-/ \sqrt{-p_-^2}, \\
&&(a_1)_{\mu} = \varepsilon_{\mu \nu \kappa \sigma}a_0^{\nu}a_3^{\kappa}a_2^{\sigma},\;
(a_{2})_{\mu} = \varepsilon_{\mu \nu \kappa \sigma}p_1^{\nu}p_2^{\,\kappa}q_1^{\sigma}/\rho \,. \nn
\ea
Here, $p_{\pm}=p_2 \pm p_1$, and $\rho$ is determined from the normalization conditions
$a_{1}^{2} = a_{2}^{2} = a_{3}^{2}=-a_{0}^{2}=-1$, $a_{\delta_e}=a_1+i\delta_e a_2$,
$a_{\delta_e}^{\ast}=a_1-i\delta_e a_2$, $\delta_e=\pm 1$, $\tau_e=-p_-^2/4m_e^2$.

For the leptonic tensor man can be written expression similar to (\ref{WmunuDSB}):
\ba
\label{LmunuDSB}
&&L_{\mu \nu}^{\delta_{e_1} \delta_{e_2}}=4m_e^2\left [\frac{(1 + \delta_{e_1}
\delta_{e_2})}{2}\, (a_0)_{\mu} (a_0)_{\nu}+ \right. \nn \\
&&~~~~~\left.+\frac{(1 - \delta_{e_1} \delta_{e_2})}{2} \, \tau_e \;
(a_{\delta_e})_{\mu} (a^{\ast}_{\delta_e})_{\nu}\right ]\,.
\ea

Let us consider the question of the polarization state of a virtual photon
with 4-momentum $q=p_{1}-p_{2}=q_2-q_1$ which is exchanged between the
electron and proton in the reaction $ep \to ep$. Using the vectors of the orthonormalized
basis $a_{A}$ (\ref{OBVe}) which satisfy the completeness condition
\ba
{a_0}_{\mu} \, {a_0}_{\nu} - {a_1}_{\mu} \, {a_1}_{\nu} - {a_2}_{\mu} \, {a_2}_{\nu} - {a_3}_{\mu} \, {a_3}_{\nu}
= g_{\mu \nu} \; ,
\label{7.4}
\ea
we define the polarization vectors for the virtual photon with 4-momentum $q$ as
\ba
\label{7.5}
e_{1}& =& \frac{(a_{1}q_1)\, a_{0} -(a_0 q_1)\, a_1} {\sqrt{(a_{3}q_{1})^2 + q_{1}^2} } , \;
\;\; e_{2} = a_{2} \,, \\ 
e_{3} &=& \frac{ q_1 + (a_{3} q_1)\, a_{3}} {\sqrt{(a_{3}q_{1})^2 + q_{1}^2} } \; ,
\;\; (e_{2})_{\mu} = \varepsilon_{\mu \nu \kappa \sigma}p_1^{\nu}p_2^{\,\kappa}q_1^{\sigma}/\rho \,,\nn
\ea
where $e_{1}$ and $e_{2}$ are the transverse polarization vectors, $e_{3}$ is
the longitudinal polarization vector, and
\ba
&& ~~~~~~~~~~~~~~~~~~~ \rho^2 = (a_{1}q_{1})^2 =  \\
&& ~~ = [\, 2(p_{1}p_{2}) (p_{1}q_{1}) (p_{2}q_{1}) - M^2 ((p_{1}p_{2})^2 - m_e^4) - \nn \\
&& ~~ - \,m_e^2\, ((p_{1}q_{1})^2 + (p_{2}q_{1})^2)\,]/ ( (p_{1}p_{2})^2 - m_e^4) \nn\,.
\ea
It is easily verified that the 4-vectors $e_{i} \, (i=1, 2, 3)$ are orthogonal
to each other ($e_{i} e_{j} = 0, \, i \neq j$), and also that $e_{i}q
= e_{i}a_{3}= 0$ and $e_{1}^2 = e_{2}^2 = - e_{3}^2 = -1$.

In the rest frame of the initial proton ($q_{1}= (M,0,0,0))$ the 4-vectors $e_{i}$
have the form:
\ba
&& e_{1} = (0,1,0,0), \; e_{2} = (0,0,1,0),\nn  \\
&& e_{3} = \frac{1}{\sqrt{-q^2}} (\mid \vecc q \mid, q_{0} \vecc n_{3}) \; .
\label{7.6}
\ea
Here $\vecc n_{3}$ is a unit 3-vector directed along $\vecc q \;(\vecc n_{3}^{~2}
=1)$, and $q_{0}$ is the time component of the 4-vector $q=(q_{0}, \vecc q)$.

The vectors $e_{1}, e_{2}, e_{3}$, and $a_{3}$ are orthogonal to one another and also satisfy
the completeness condition
\ba
{e_3}_{\mu} \, {e_3}_{\nu} - {e_1}_{\mu} \, {e_1}_{\nu} - {e_2}_{\mu} \, {e_2}_{\nu} - {a_3}_{\mu} \, {a_3}_{\nu}
= g_{\mu \nu} \; ,
\label{7.7}
\ea
which makes it possible to express $a_{0}$ and $a_{1}$ in terms of $e_{1}$ and $e_{3}$ as
\ba
&& a_{1}=\alpha e_{3} - \beta e_{1} , \; a_{0} = \beta e_{3} - \alpha e_{1}, \; \beta^2=1 + \alpha^2 ,\nn \\
\label{7.8}
&& \alpha = e_{3} a_{1} = a_{0} e_{1} = \frac { a_{1}q_{1}}{\sqrt{(a_{3}q_{1})^2  + q_{1}^2} }\; , \\
&& \beta = e_{1} a_{1} = e_{3} a_{0} = \frac {a_{0} q_{1}} {\sqrt{(a_{3}q_{1})^2 + q_{1}^2} } \; .
\label{7.9}
\ea
The matrix elements of the electron current (\ref{Jel-pp}), (\ref{Jel-pm})
in terms of the 4-vectors $e_{i}$ (\ref{7.5}) can be represented as
\ba
\label{7.11}
(J^{\delta_e,\delta_e}_{e} )_{\mu}&=& 2 m_e (\beta e_{3} - \alpha e_{1})_{\mu} \ , ~~~\\
( J^{-\delta_e,\delta_e}_{e} )_{\mu} & = & -2 m_e \delta_e \sqrt{\tau_e} \,(a_{\delta_e})_{\mu}, ~~~\\
(a_{\delta_e})_{\mu}& = &(\alpha e_{3} - \beta e_{1} + i \delta_e e_{2})_{\mu}.
\label{7.11a}
\ea
Therefore, for the transition without electron spin-flip $(J^{\delta_e,\delta_e}_{e})$
the virtual-photon polarization vector is a superposition of the longitudinal
($\beta e_{3}$) and transverse linear ($-\alpha e_{1}$) polarizations, while
for the transition with spin-flip $(J^{-\delta_e,\delta_e}_{e})$ it is a superposition
of the longitudinal ($\alpha e_{3}$) and the transverse elliptical ($e_{\delta_e}$)
polarizations
\be
e_{\delta_e} \equiv  -\beta e_{1} + i \delta_e e_{2} =(0, \vecc e_{\delta_e}), \,\beta^2 \geq  1 \,.
\ee
The state of a photon with elliptical polarization vector
$e_{\delta_e} =(0, \vecc e_{\delta_e})$ will have degree of linear polarization (equal to
the ratio of the difference and sum of the squared ellipse semiaxes:
\ba
\label{7.12}
&& \kappa_{\gamma} = \frac{\beta^2 - 1}{\beta^2 + 1} = \frac{\alpha^2}{\alpha^2 + 2} \; ,\\
&& \kappa_{\gamma}^{-1}=1 + \frac{2}{\alpha^2}=1+2\, \frac{\,(a_{3}q_{1})^2  + q_{1}^2\,}{(a_{1}q_{1})^2}\;.
\ea
Inverting relation in Eq. (\ref{7.12}) we obtain:
\be
\beta^2 =\frac { 1 + \kappa_{\gamma}}{1 - \kappa_{\gamma}} \; \; , \;
\alpha^2= \frac{2 \kappa_{\gamma}} {1 - \kappa_{\gamma}} \; \; .\nn
\label{7.13}
\ee
Let us find the squared moduli of the vectors $\vecc e_{\delta}$ and $\vecc a_{\delta}$:
\ba
\label{7.14}
&& \mid \vecc a_{\delta_e} \mid^2 =  (1 + \kappa_{L}) \,\mid \vecc e_{\delta_e} \mid^2 \,, \\
&& \mid \vecc e_{\delta_e} \mid^2 = 1 + \beta^2 = \frac {2}{1 - \kappa_{\gamma}} \; ,\;  \\
&& \kappa_{L} = \kappa_{\gamma} \, \vecc e_{3}^{~2} \,, 
\; \vecc e_{3}^{~2} = \frac{ q_{0}^2 }{(-q^2)} \; .
\label{7.15}
\ea
Let us introduce the normalized vectors $\vecc e_{\delta}'$ and $\vecc a_{\delta}'$:
\ba
\label{7.16}
&& \vecc e_{\delta_e}' = \frac {\vecc e_{\delta_e}}{\sqrt{1 + \beta^2}}
= \sqrt {\frac {1 - \kappa_{\gamma}}{2}} \; \vecc e_{\delta_e} \; , \;   \\
\label{7.16a}
&& \vecc a_{\delta_e}' = \frac {\vecc a_{\delta_e}}{\sqrt{1 + \beta^2}} =
\sqrt {\frac {1-\kappa_{\gamma}}{2}} \; \vecc a_{\delta_e} \; , \\
&& |\vecc e_{\delta_e}^{'}|^2 = 1 \; , \; |\vecc a_{\delta_e}^{'} |^2
= 1 + \kappa_{\gamma} \vecc e_{3}^{~2} = 1 + \kappa_{L} \; .
\label{7.17}
\ea
Therefore, the elliptical-polarization vector $\vecc e_{\delta_e}$ of a virtual
photon can be normalized to unity ($|\vecc e_{\delta_e}'|^2 = 1$), but the
presence of a longitudinal polarization makes this normalization impossible
for the total vector $\vecc a_{\delta_e} '$ simultaneously. The quantity
$\kappa_{L}$ (\ref{7.15}) corresponding to the inequality $|\vecc a_{\delta_e} '
|^2 = 1 + \kappa_{L} \neq 1$ has the meaning of the degree of longitudinal
polarization of a virtual photon emitted in a transition with electron spin flip.

\subsection{Ultrarelativistic, massless case}

In the ultrarelativistic limit, when the electron mass can be neglected,
the matrix elements of the electron current (\ref{Jel-pp}), (\ref{7.11}) without
spin-flip are vanished. In this case all the polarization characteristics of a virtual photon
are determined by the vector (\ref{7.11a}). In this (massless) case
the quantities $\kappa_{\gamma}$ (\ref{7.12}) and $\kappa_{L}$ (\ref{7.15})
will be interpreted as the total degrees of linear and longitudinal polarization of the virtual photon.
For this (massless) case we have:
\ba
\label{7.18}
&& (a_{3}q_{1})^2 + q_{1}^2 = - m^2\;\frac {\, \vecc q^{2}}{q^2} \; , \\
&& (a_{1} q_{1}) ^2 = m^2 \; \cot^2 (\vartheta_e /2) \; ,\\
&& ~~\kappa_{\gamma}^{-1} = 1 - 2 \; \frac {\,\vecc q^{2}} {q^2} \; \tan^2 (\vartheta_e /2) \; ,
\label{7.19}
\ea
where $\vartheta_e$ is the angle between the vectors $\vecc p_{1}$ and
$\vecc p_{2}$. Using relations
\ba
\vecc q^2= 4 \,m^2\,\tau(1+\tau),\, q^2=-4\,m^2\,\tau, \, q_0=2m \tau\, ,
\ea
expression (\ref{7.19}) can be rewritten in another form
\ba
\kappa_{\gamma}^{-1} = 1 + 2 \, (1+\tau) \, \tan^2(\vartheta_e /2) \; ,
\ea
that coincides with the result for the quantity $\varepsilon^{-1}$ in Rosenbluth
formula (\ref{Ros}). For the degree of the virtual photon longitudinal polarization
we have
\ba
\kappa_L=\kappa_{\gamma}\, \tau \,.
\ea

Note the vector $\vecc a'_{\delta_e}$ (\ref{7.16a}) can also be written as
\ba
\vecc a'_{\delta_e} = \sqrt{\kappa_{L}}  \vecc n_{3} - \sqrt {\frac{1 +\kappa_{\gamma}}{2}}
\vecc e_{1} + i \delta_e  \sqrt {\frac {1 - \kappa_{\gamma}}{2}} \vecc e_{2} \, ,
\label{7.20}
\ea
which makes it easy to construct the polarization density matrix for a virtual
photon in the massless limit both in the polarized case, which for massless
particles is helical polarization, and in the unpolarized case; see \cite{Rekalo68,GLev}.

\end{document}